\documentclass[letterpaper,twocolumn,10pt]{article}
\usepackage{usenix,epsfig,endnotes,hyperref}
\begin{document}

\date{}

\title{\Large \bf Trustware: A Device-based Protocol for Verifying Client Legitimacy}


\author{
{\rm Ben Doyle, Patrick Korth, Kyle Nekritz, Zane Salem}\\
The University of Michigan
} 

\maketitle

\thispagestyle{empty}

\subsection*{Abstract}
Online services commonly attempt to verify the legitimacy of users with CAPTCHAs. However, CAPTCHAs are annoying for users, often difficult for users to solve, and can be defeated using cheap labor or, increasingly, with improved algorithms. We propose a new protocol for clients to prove their legitimacy, allowing the client's devices to vouch for the client. The client's devices, and those in close proximity, provide a one-time passcode that is verified by the device manufacturer. This verification proves that the client has physical access to expensive and trusted devices, vouching for the client's legitimacy.


\section{Introduction and Background}
Many online services provide functionality that is intended for use by humans, but can be abused for profit by automated robots. For instance, robots can sign up for free email accounts (GMail, Yahoo! Mail, etc.) and send spam. Service providers would like to prevent automated bots from signing up for these services, and ensure that only a real human can sign up. Prevention of automated crawling and brute-force password guessing attempts is also desired.\\

Currently, CAPTCHAs are used to attempt to differentiate between humans and automated bots. A CAPTCHA is a puzzle presented to users that is believed to be easy for a human to solve, but very difficult for a computer to algorithmically solve. The user must correctly solve the puzzle to prove to the service that they are a human. The most common CAPTCHAs used today are images containing distorted text that users must transcribe. However, with improvements in OCR (optical character recognition) algorithms, these puzzles have become easier for computers to solve algorithmically. Additionally, adding more distortion to the images to counteract the OCR improvements have made many CAPTCHA puzzles very difficult for humans to solve. These CAPTCHA puzzles also have large accessibility problems (for users with vision impairment, etc.).\\


However, all puzzle-based CAPTCHAs have a fundamental problem: even if a perfect puzzle can be created — one that can be consistently easily solved by humans, but can never be algorithmically solved by a bot — attackers can still hire humans to solve the puzzles for them. API’s have been developed that use third-world labor in order to do this for fractions of a penny a piece\footnote{https://2captcha.com/}. If the value that can be gained by the attacker (and lost by the service provider) is of any significance, the CAPTCHA is fairly useless.\\

We propose Trustware: a novel method of verifying legitimate clients. Trustware is not a CAPTCHA, but instead is built off of the observation that when a legitimate user is signing up for a service, (s)he is likely doing so from a variety of different electronic devices: a laptop or desktop computer, a smartphone, a tablet, etc. All of these devices are expensive pieces of hardware, likely costing at least \$100. Increasingly, users are also likely to be in close proximity to a number of Internet of Things sensor devices like FitBits or smart watches.\\

When an online service requests that a client proves their legitimacy using Trustware, the client gathers one-time password tokens from nearby devices. The one-time password tokens are generated using a secret placed on each device by the device's manufacturer. This allows the manufacturer to verify that the device is in the presence of the client, and confirm as much to the online service.


\section{Related Work}

CAPTCHAs \cite{captcha} first became popular in the early 2000's as a security mechanism against malicious automated robots. In the years since, both algorithms to defeat CAPTCHAs and CAPTCHAs themselves have developed. ReCAPTCHA \cite{von2008recaptcha}, developed in 2008, uses scanned words that OCR algorithms are unable to decipher as CAPTCHAS puzzles.\\

Google's latest attempt at CAPTCHA, NoCAPTCHA \cite{NOCAPTCHA}, is of particular interest. NoCAPTCHA uses information Google has collected on users through its array of services in addition to a traditional CAPTCHA to determine if a user is legitimate. If enough information is present on the client, the CAPTCHA puzzle may even be skipped. However, this approach requires a large amount of data available on users, which is often not practical, and comes with many privacy concerns.\\

Prior work has also been done to verify a client's possession of devices. Two-factor authentication keys use this principle. When a two-factor key is manufactured, a secret is implanted into it, and one-time passwords are generated from it, for confirmation from an online service. In particular, U2F\footnote{https://www.yubico.com/applications/fido/} aims to create a standardized protocol for verification of a client's possession of a keychain device. However, these approaches focus on confirming the identify of a specific user, not on simply deciding whether a client has legitimate intentions.


\section{Architecture}

\begin{figure*}
\begin{center}
\includegraphics[width=\textwidth]{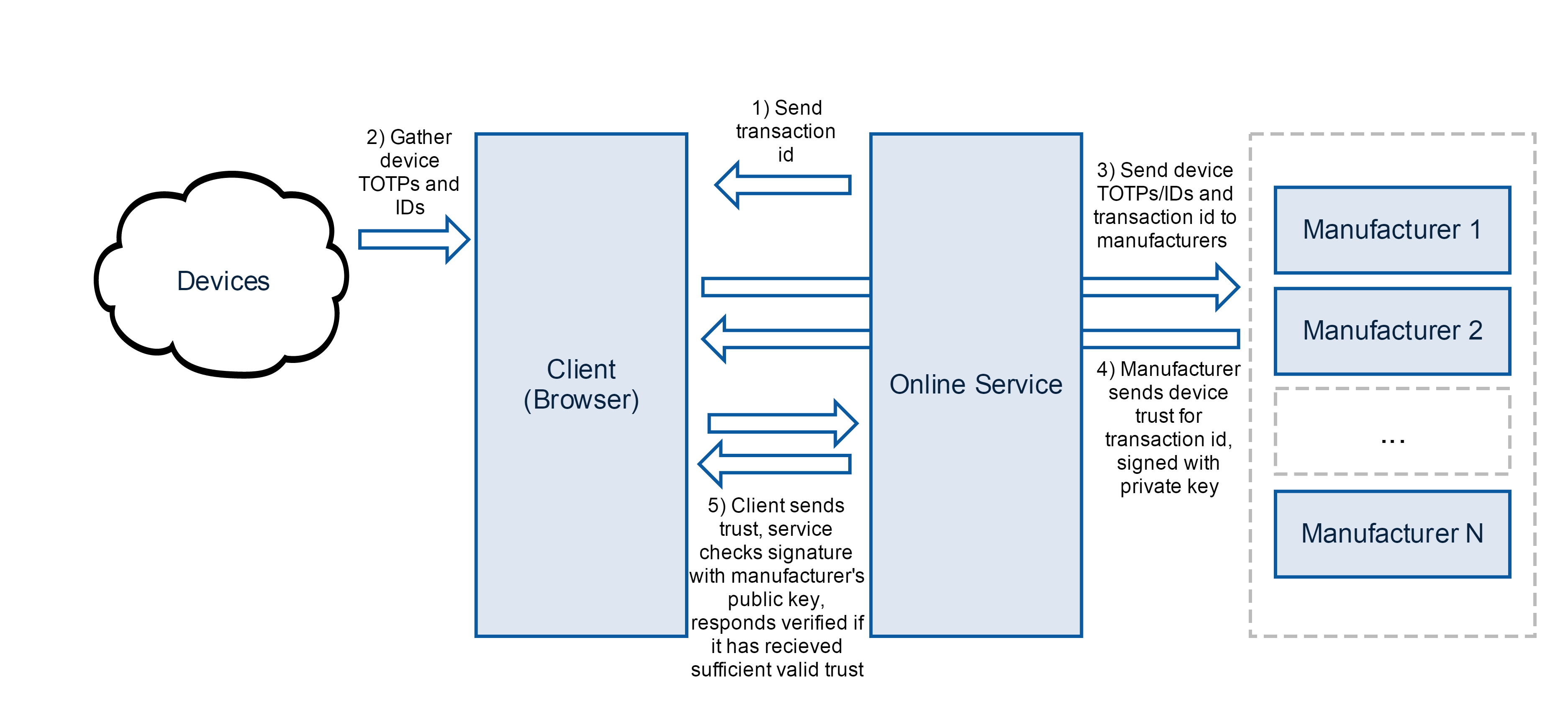}
\caption{Trustware Protocol}
\label{fig:correctprotocoldiagram}
\end{center}
\end{figure*}

We have designed a communication protocol that satisfies the goals set out for the project while maintaining as much user privacy and convenience as possible. Furthermore, our protocol ensures that various attacks and exploitations are not viable. We first give an overview of the components that make the protocol possible, then step through our proposed protocol, and finally, describe a practical implementation of the protocol.

\subsection{System Components}

There are four prerequisites to be set up for Trustware authentication: the device manufacturer authentication server, hardware and software support on client-side devices, website-side Trustware software, and client browser support. \\

For a particular brand of electronic device to be supported by Trustware, the company that manufacturers the device must run an authentication server in order to authenticate its own devices. This is a simple web service that tracks produced devices and associated Trustware information. When a device is manufactured, it is assigned a unique ID and secret key. The server maintains a database of each device ID, secret, and usage information. The server responds to device-trust requests from clients and allows the company to register new devices as they are manufactured. Devices must also run software that can communicate with the client-side web browser. This will vary depending on the nature of the device; exact protocol specifications are described in the next section. \\

Each device will be manufactured with three hard-coded values: ID, secret key, and company authentication server URL. Unique IDs are randomly assigned by the company that manufactures the device, as is the base32-encoded 16-character secret key. These two values are ideally as difficult as possible to extract or alter on a device. The device will also contain the URL of the authentication server run by the manufacturer. \\

In order for a website to verify legitimate users via Trustware, it must embed Trustware code in its verification process. When a user clicks "Submit" (or whatever action needs to be verified), the embedded coordinates with the client's web browser to verify the legitimacy of the client. The software then returns whether or not it believes the client is a legitimate user. \\

In order to coordinate communications between devices, manufacturers, and the web service, the client must be using a web browser that supports Trustware. The browser detects Trustware-enabled websites, then communicates with the client's devices and associated manufacturers to send authenticated trust levels to the web service. \\

\subsection{Protocol}

We begin when a webpage requires verification that users are legitimate users. This could be a multitude of reasons: email sign-up, auction bidding, forum posting, etc. A Trustware-enabled client visits the website, and it generates a new session token for the client. This session token is stored in the HTML of the webpage in order to communicate it to the client (Figure \ref{fig:correctprotocoldiagram}, step 1). The client's web browser detects that the webpage is Trustware-enabled, grabs the token, and begins harvesting nearby device information. \\

Trustware devices near the client's computer broadcast three identifying pieces of information: the device ID, a time-based one-time-password (TOTP), and the URL of the device manufacturer authentication server. The TOTP is calculated using standard TOTP practices: the current 30-second time interval since Unix epoch is combined with the device's secret key using HMAC to generate a token that is only valid at the time it was generated, within 30 seconds \cite{TOTP}. Upon receiving a device information tuple, the web browser makes a HTTPS request with the device ID, TOTP, and client session token to the device manufacturer URL in order to authenticate (Figure \ref{fig:correctprotocoldiagram}, steps 2 and 3). \\

The device manufacturer authentication server receives the request. It retrieves the secret key for the given device ID from its device database, and computes the TOTP for the current 30-second interval. If the given TOTP and computed TOTP match (or it matches an adjacent interval's TOTP code, accounting for clock skew), the request has been verified as legitimate and the process continues. The server then takes the device authentication history and calculates the relative amount of trust for the device. This heuristic can vary by manufacturer, but in general, it has the desirable property that devices with usage behavior similar to automated bot systems are given low trust, whereas devices with human-like usage behavior are given high trust. As an example, if requests have been made to authenticate a device multiple times in the span of a few seconds, these requests are most likely coming from an automated bot and low trust should be given. After calculating device trust, and server responds to the client request with the client session token, device trust, and current time, all signed with the private key of the manufacturer (Figure \ref{fig:correctprotocoldiagram}, step 4). \\

Manufacturer responses will steadily come in to the client, which are then forwarded to the requesting webpage (Figure \ref{fig:correctprotocoldiagram}, step 5). Each response is a tuple indicating that the manufacturer vouches (or doesn't) for the trust of a given device in the client's vicinity. When the webpage receives a device trust tuple, it checks that it is coming from a verified manufacturer. The webpage has a list of Trustware-verified manufacturers and associated public keys, and verifies the tuple using this information. Incorrectly-signed tuples or unverified manufacturers are rejected. The enclosed device trust is then added to the overall trust level for the client's session. If the trust level reaches some predetermined value (chosen by the webpage), the client is deemed legitimate and allowed access to whatever resource the webpage offers. Should the client fail to provide enough verified device trust to satisfy the webpage within some predetermined length of time (chosen by the webpage), the client is assumed to be illegitimate and subsequently rejected.

\subsection{Implementation}

\begin{figure*}
\begin{center}
\includegraphics[width=\textwidth]{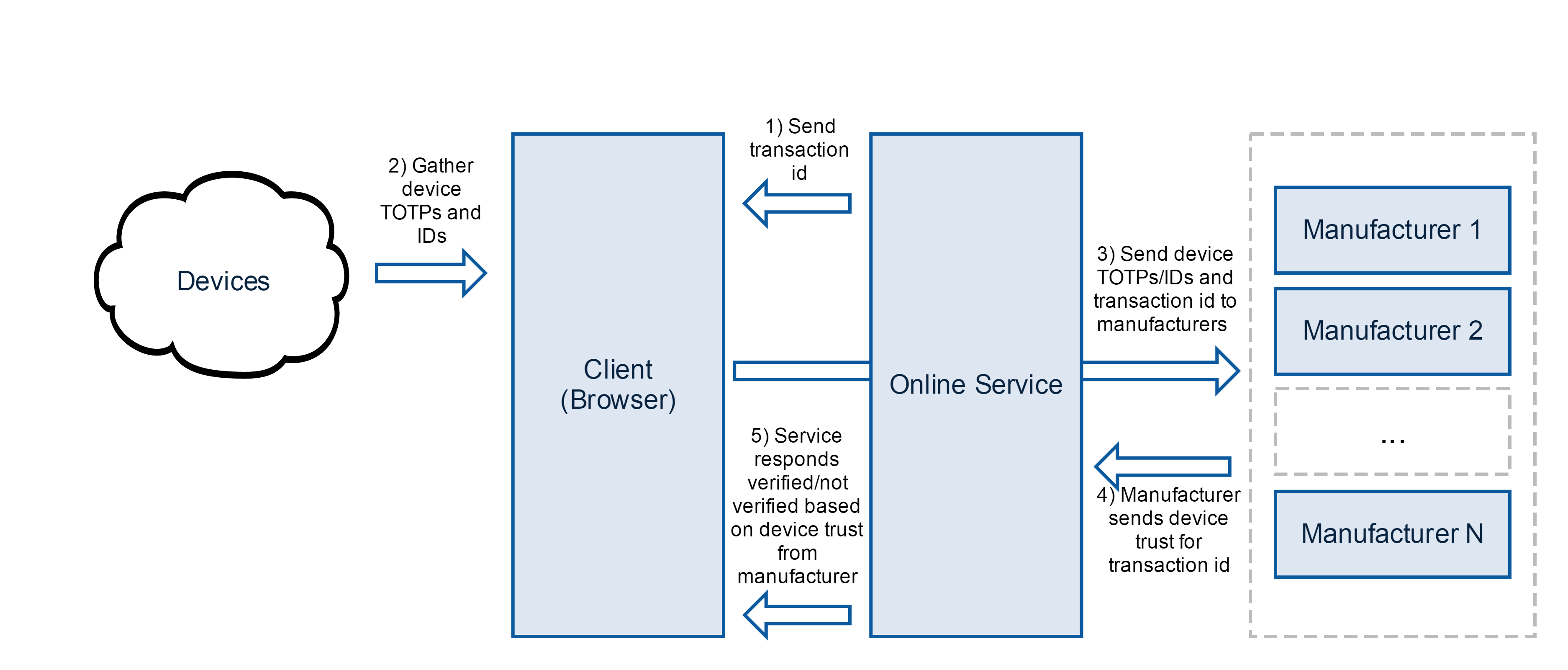}
\caption{Implementation}
\label{fig:ourimplementation}
\end{center}
\end{figure*}

We have implemented an example Trustware system in order to illustrate the protocol. The client can visit our Trustware-enabled webpage, gather device information from an Android phone and Linux desktop, verify these devices with a mock manufacturer server, then relay trust information to the webpage and gain access. There are several important differences between our implementation and proposed protocol:

\begin{itemize}
\itemsep0em
\item Device IDs and secrets are stored on the software level, rather than the hardware level
\item A web browser plugin facilitates communications, rather than the web browser having the functionality built in
\item Manufacturer trust responses are sent directly to the requesting webpage, rather than relayed through the client (compare steps 4 and 5 in Figure \ref{fig:correctprotocoldiagram} vs \ref{fig:ourimplementation})
\item Device trust is calculated using a simple rate-limiting heuristic, rather than a more intelligent measure
\end{itemize}

All of the components described below can be found online at \href{https://github.com/trustware}{https://github.com/trustware}. Each component has installation and usage instructions, making it possible to set up and use the entire Trustware system. \\

\noindent
\textbf{Android phone:} We have developed an Android application that uses Bluetooth LE to advertise its presence to Trustware. The application sends advertisement packets containing the device ID, current TOTP, and manufacturer URL. As previously noted, the device ID is stored on a software level as a compile-time constant. Our web browser plugin, below, listens for these packets when searching for nearby Trustware devices. The Android phone must have software and hardware support for Bluetooth LE; we tested the application on a Samsung Galaxy S5 running Android 5.0. \\

\noindent
\textbf{Linux driver:} This driver runs on a desktop Linux system in order to advertise its presence to Trustware. It advertises the same device tuple as the Android phone, and similarly has a software-level device ID. Advertisement information can be read by the web browser plugin, below, via /dev/otp. The driver was installed on Ubuntu 14.04 and set to run at start-up. \\

\noindent
\textbf{Web service:} We created a dummy webpage that requires Trustware authentication. The webpage allows users to create an account on an imaginary service, such as an email provider. It communicates with the client as specified in the Trustware protocol, but as previously noted, is sent device trust information by the manufacturer directly rather than by through the client. If the webpage is not sent sufficient device trust within a few seconds, it rejects the client's request. We wrote a simple script in Javascript that website developers can easily embed in their own webpages. \\

\noindent
\textbf{Google Chrome extension:} Rather than modifying a web browser, we elected to write a plugin to carry out the Trustware protocol. We wrote an extension for Google Chrome that watches for Trustware requests from webpages, then relays device trust information to the corresponding manufacturer servers. It scans the HTML of visited webpages for a custom attribute tag that identifies Trustware and contains the session token. If a webpage requested Trustware verification, the plugin scans Bluetooth LE advertisement packets (Android phone) and /dev/otp (Linux driver) for device information. Device tuples are then sent to the mock manufacturer server in order to be verified. \\

\noindent
\textbf{Mock manufacturer:} We set up a manufacturer authentication server to serve the aforementioned Trustware devices. The server maintains a database of device ID, secret, and use count tuples. When an authentication request is received, the server follows the Trustware protocol in order to verify the device. Device trust is calculated as an inverse of the number of uses and rate-limited; i.e., device trust decreases as the number of authentication requests increases, and a device cannot authentication twice in ten seconds. This is a simplistic trust heuristic simply for demonstration purposes, and is improper for actual deployment. After calculating device trust, a signed tuple of session token and device trust is sent to the web service. \\

\section{Evaluation}
Trustware aims to move the necessary verification work from the client's perspective to browser, online service providers, and device manufacturers. In our evaluation we consider the adoption cost for each stakeholder: end users, web browsers, online services, and device manufacturers. We also discuss the usefulness as well as limitations of our protocol in practice.

\subsection{Costs}
\textbf{User:} Users must be surrounded by or use machines that have Trustware support to use Trustware verified services. Each message sent and received in the Trustware protocol involves \textless 1 KB of data, clearly within the constraints of a single network packet for each trip made. A client is required to establish additional sessions with a manufacturer service. More implicitly is the cost for total user privacy. Because the online service must verify trust vouches using a specific manufacturer's public key, the manufacturers, or device brands, that surround a user are announced to the online service in use. \\

\noindent
\textbf{Browser:} A compatible web browser is required to retrieve (TOTP, ID, Manufacturer) tuples on Trustware form submissions through the device in use as well as in range Trustware Bluetooth LE advertisement packets.\\

\noindent
\textbf{Web Service:} A web service that wishes to integrate Trustware verification must make frontend and backend modifications to their site. The complexity of this completely rides on ease of its framework. Though we've implemented such a framework, one as robust and finalized as Google's CAPTCHA service would be ideal for adoption. \\

\noindent
\textbf{Manufacturer:} Each manufacturer will have many responsibilities. First, a secret Trustware key needs to be added into devices at manufacturing time (ideally implanted in hard to extract Trusted Platform Modules). Second, they must store and accept requests for such OTP being sent for that particular device ID. \\

\subsection{Value Added}
  Once the Trustware automated pipeline is set up, verifying client legitimacy can be done without manual work from any participant while providing the benefits of CAPTCHA. An automated legitimacy checking process is also faster than an average CAPTCHA solver \cite{captchaspeed}. \\

  
\subsection{Limitations}

The goals of the Trustware protocol force some limit on what it is meant to accomplish. To reiterate, Trustware does not prove that someone is human. It's merely a reputation based rate limiting tool that is apathetic toward someone's true identity.\\

Because of no client-side checks, Trustware also cannot defend against a compromised machines. For instance, a bot net could exploit a user's device and the device's surrounding them for trust. Trustware is also currently susceptible to what we've deemed "trust-eating" and "trust-mining". Imagine an adversary that has set up in a location surrounded by many Trustware broadcasting devices. By collecting surrounding device's tuples (OTP, ID, manufacturer) and requesting verification from the manufacturer over and over (though limited by the manufacturer service's timeout "recharge" for that device), this adversary could use the collected trust for services or just throw it away, unnecessarily tampering with other's reputations. More impressively, adversaries could collude in their efforts collecting trust for prespecified tokens outside their vicinity. A trust mining business or API could even be created to collect trust to assign to a given session token. This attack's efforts are however limited to a small timeframe (on our current recommended trust heuristic 30 seconds). Stockpiled trust would effectively expire and is only valid for a specific token. We've concluded "trust-mining" attacks cost more than remote CAPTCHA solvers because these devices are more costly puzzle solvers. \\ 

Due to the device reputation based system, shared resources, such as a library or university computing center, is also arguably bound not to have trust values that reflect the fact that it is shared. However, these shared devices arguably have trust levels that are reflect their use.  \\

Lastly, because of Trustware's vicinity requirement, users who remotely access a machine not within Bluetooth range also will not have adequately reflected trust. One possible remedy is to remotely transmit trust to said machine, similar to our "trust-mining" attack effort.


\section{Future Work}

We have demonstrated a proof-of-concept implementation of the Trustware protocol. In addition to fine-tuning the trust heuristic and timeout period to defend against mining, future work includes securing the participation of hardware device manufacturers and increasing the privacy awareness of the implementations. 

\subsection{Manufacturer Participation}

The participation of the device manufacturers is a crucial for the success of the protocol. In order for Trustware to function it was assumed manufacturers had an API to respond with trust levels for a given device ID. In reality this requires hardware companies to pay for a dedicated server, as well as to securely encode the device's secrets and API endpoint URLs at manufacture time. Although such costs are negligible a reasonable question to ask is why a hardware manufacturer would be willing bear them to support Trustware.\\

Many large players in the device production industry also have a vested interest in trying to solve the problem of verifying legitimate clients. One such example is Google, which is not necessarily a traditional hardware manufacturer, but has a wide variety of online services for which they only want legitimate clients to be able to use. If a large player like Google starting using the protocol as a standard they could use their leverage over the Android ecosystem to compel many device manufacturers to follow suit.\\

Google demonstrated their desire to tackle this problem by releasing their own solution in "No CAPTCHA reCAPTCHA" which utilizes the vast amount of data they have on individual clients to determine their legitimacy \cite{NOCAPTCHA}. However a problem with Google solution is many developers don't want to provide Google with more data for privacy and competitive reasons so they may be reluctant to opt-in and instead continue to offload the work of proving client legitimacy to their users. Both issues of preventing tedious work for the user and respecting their privacy could be addressed through adopting Trustware. \\

Alternatively, a more ambitious line of exploratory work could be the use of a carefully designed decentralized "trust token" on the Ethereum  \cite{wood2014ethereum} network to take the place of a manufacturer server.

\subsection{Privacy Improvements}

User privacy is promisingly becoming a more important consideration. Making the manufacturer blind to what services are using the protocol was one way to increase privacy awareness. We also propose using rotating device IDs in the manufacturer's implementation.\\

Users would likely be hesitant to supply a static unique ID as currently implemented, as it provides an easy way of tracking their activity. Fortunately, the specifications for Bluetooth Low Energy \cite{BLESecurity} are very privacy minded and allow for developers to use rotating IDs to frequently change the private addresses of devices to avoid such tracking.\\

This increased privacy awareness and the decentralized nature of the Trustware protocol will hopefully make it a more compelling alternative to current no-CAPTCHA methods for verifying client legitimacy.\\

The entire Trustware protocol has a realm of other possible use cases, one of which is a reputation based crowd sourcing for things such as location and other spoof-able things. We believe there is more interesting work to be done here.


\section{Conclusion}

Trustware provides a unique verification system that approves legitimate clients while rejecting malicious ones. Work still needs to be done developing a powerful standard for trust heuristic and the trade-off between absolute convenience and privacy/protection. Yet, we still believe that there is an incentive for someone in the online services and hardware space (Google, Apple, etc.) to take notice. With the increasing prevalence of devices using Bluetooth LE in the near future that Trustware can piggyback on, this is well within the realm of possibility.


{\footnotesize \bibliographystyle{acm}
\bibliography{usenixTemplate.bib}}

\end{document}